\documentclass[%
reprint,
superscriptaddress,
 amsmath,amssymb,
 aps,
 pra,
]{revtex4-1}

\usepackage{graphicx}% Include figure files
\usepackage{dcolumn}% Align table columns on decimal point
\usepackage{bm}

\usepackage[utf8]{inputenc} 

\usepackage{multirow}
\usepackage[warn]{mathtext}
\usepackage{amsfonts}
\usepackage{amsmath}
\usepackage{amssymb}
\usepackage{braket}
\usepackage{bbold}
\usepackage{textcomp} 
\usepackage{indentfirst} 
\usepackage{amsmath} 
\usepackage{graphicx}

\DeclareGraphicsExtensions{.pdf,.png,.jpg}
\usepackage{hyperref}

\usepackage{algpseudocode}

\renewcommand{\bibname}{References}

\usepackage[nottoc]{tocbibind}

\usepackage{xcolor}
\hypersetup{
    colorlinks,
    linkcolor={red!50!black},
    citecolor={blue!50!black},
    urlcolor={blue!80!black}
}

\newcommand{\T}{\mathcal{T}}
\renewcommand{\P}{\mathcal{P}}



\begin{document}

    \title{Demonstration of a parity–time symmetry breaking phase transition \\ using superconducting and trapped-ion qutrits}

    \author{Alena~S.~Kazmina}
    \affiliation{Russian Quantum Center, Skolkovo, Moscow 121205, Russia}
    \affiliation{National University of Science and Technology MISIS, Moscow 119049, Russia}
    \affiliation{Moscow Institute of Physics and Technology, Dolgoprudny 141700, Russia}

    \author{Ilia~V.~Zalivako}
    \affiliation{Russian Quantum Center, Skolkovo, Moscow 121205, Russia}
    \affiliation{P.N. Lebedev Physical Institute of the Russian Academy of Sciences, Moscow 119991, Russia}

    \author{Alexander~S.~Borisenko}
    \affiliation{Russian Quantum Center, Skolkovo, Moscow 121205, Russia}
    \affiliation{P.N. Lebedev Physical Institute of the Russian Academy of Sciences, Moscow 119991, Russia}
    
    \author{Nikita~A.~Nemkov}
    \email{nnemkov@gmail.com}
    \affiliation{Russian Quantum Center, Skolkovo, Moscow 121205, Russia}
    \affiliation{National University of Science and Technology MISIS, Moscow 119049, Russia}

    \author{Anastasiia~S.~Nikolaeva} 
    \affiliation{Russian Quantum Center, Skolkovo, Moscow 121205, Russia}
    \affiliation{National University of Science and Technology MISIS, Moscow 119049, Russia}

    \author{Ilya~A.~Simakov}
    \affiliation{Russian Quantum Center, Skolkovo, Moscow 121205, Russia}
    \affiliation{National University of Science and Technology MISIS, Moscow 119049, Russia}
    \affiliation{Moscow Institute of Physics and Technology, Dolgoprudny 141700, Russia}

    \author{Arina~V.~Kuznetsova}
    \affiliation{Russian Quantum Center, Skolkovo, Moscow 121205, Russia}
    \affiliation{National University of Science and Technology MISIS, Moscow 119049, Russia}
    \affiliation{Moscow Institute of Physics and Technology, Dolgoprudny 141700, Russia}

    \author{Elena~Yu.~Egorova}
    \affiliation{Russian Quantum Center, Skolkovo, Moscow 121205, Russia}
    \affiliation{National University of Science and Technology MISIS, Moscow 119049, Russia}
    \affiliation{Moscow Institute of Physics and Technology,  Dolgoprudny 141700, Russia}

    \author{Kristina~P.~Galstyan}
    \affiliation{Russian Quantum Center, Skolkovo, Moscow 121205, Russia}
    \affiliation{P.N. Lebedev Physical Institute of the Russian Academy of Sciences, Moscow 119991, Russia}

    \author{Nikita~V.~Semenin}
    \affiliation{Russian Quantum Center, Skolkovo, Moscow 121205, Russia}
    \affiliation{P.N. Lebedev Physical Institute of the Russian Academy of Sciences, Moscow 119991, Russia}

    \author{Andrey~E.~Korolkov}
    \affiliation{Russian Quantum Center, Skolkovo, Moscow 121205, Russia}
    \affiliation{P.N. Lebedev Physical Institute of the Russian Academy of Sciences, Moscow 119991, Russia}
    
    \author{Ilya~N.~Moskalenko}
    \altaffiliation{Present Address: Department of Applied Physics, Aalto University, Espoo, Finland}
    \affiliation{National University of Science and Technology MISIS, Moscow 119049, Russia}

    \author{Nikolay~N.~Abramov}
    \affiliation{National University of Science and Technology MISIS, Moscow 119049, Russia}

    \author{Ilya S. Besedin}
    \altaffiliation{Present address: Department of Physics, ETH Zurich, Zurich, Switzerland}
    \affiliation{National University of Science and Technology MISIS, Moscow 119049, Russia}

    \author{Daria~A.~Kalacheva}
    \affiliation{Skolkovo Institute of Science and Technology, Skolkovo Innovation Center, Moscow 121205, Russia}
    \affiliation{Moscow Institute of Physics and Technology, Dolgoprudny 141700, Russia}
    \affiliation{National University of Science and Technology MISIS, Moscow 119049, Russia}

    \author{Viktor~B.~Lubsanov}
     \affiliation{Moscow Institute of Physics and Technology, Dolgoprudny 141700, Russia}

    \author{Aleksey~N.~Bolgar}
     \affiliation{Moscow Institute of Physics and Technology, Dolgoprudny 141700, Russia}
     \affiliation{Russian Quantum Center, Skolkovo, Moscow 121205, Russia}

    \author{Evgeniy~O.~Kiktenko}
    \affiliation{Russian Quantum Center, Skolkovo, Moscow 121205, Russia}
    \affiliation{National University of Science and Technology MISIS, Moscow 119049, Russia}

    \author{Ksenia~Yu.~Khabarova}
    \affiliation{P.N. Lebedev Physical Institute of the Russian Academy of Sciences, Moscow 119991, Russia}
    \affiliation{Russian Quantum Center, Skolkovo, Moscow 121205, Russia}

    \author{Alexey~Galda}
    \affiliation{James Franck Institute, University of Chicago, Chicago, IL 60637, USA}

    \author{Ilya~A.~Semerikov}
    \affiliation{Russian Quantum Center, Skolkovo, Moscow 121205, Russia}
    \affiliation{P.N. Lebedev Physical Institute of the Russian Academy of Sciences, Moscow 119991, Russia}
    
    \author{Nikolay~N.~Kolachevsky}
    \affiliation{P.N. Lebedev Physical Institute of the Russian Academy of Sciences, Moscow 119991, Russia}
    \affiliation{Russian Quantum Center, Skolkovo, Moscow 121205, Russia}

    \author{Nataliya~Maleeva}
    \affiliation{National University of Science and Technology MISIS, Moscow 119049, Russia}
    
    \author{Aleksey~K.~Fedorov}
    \email{akf@rqc.ru}
    \affiliation{Russian Quantum Center, Skolkovo, Moscow 121205, Russia}
    \affiliation{National University of Science and Technology MISIS, Moscow 119049, Russia}
    \affiliation{P.N. Lebedev Physical Institute of the Russian Academy of Sciences, Moscow 119991, Russia}

\date{\today}
 
	\begin{abstract}
	Scalable quantum computers hold the promise to solve hard computational problems, such as prime factorization, combinatorial optimization, simulation of many-body physics, and quantum chemistry. While being key to understanding many real-world phenomena, simulation of non-conservative quantum dynamics presents a challenge for unitary quantum computation. In this work, we focus on simulating non-unitary parity-time symmetric systems, which exhibit a distinctive symmetry-breaking phase transition as well as other unique features that have no counterpart in closed systems. We show that a qutrit, a three-level quantum system, is capable of realizing this non-equilibrium phase transition. By using two physical platforms -- an array of trapped ions and a superconducting transmon -- and by controlling their three energy levels in a digital manner, we experimentally simulate the parity–time symmetry-breaking phase transition. Our results indicate the potential advantage of multi-level (qudit) processors in simulating physical effects, where additional accessible levels can play the role of a controlled environment. 
    \end{abstract}
	
\maketitle
		
\section{Introduction} \label{sec: intro}
	
Quantum simulation is one of the key prospective applications for quantum computing~\cite{Lloyd1996,Nori2009,Cirac2012,Nori2014,Fedorov2022, Hoefler2023}. It uses a well-controlled quantum device to replicate the behavior of the system of interest. There are two main approaches to quantum simulation. One is the analog quantum simulation, which relies on special-purpose quantum systems and can be based on a variety of platforms including superconducting transmons~\cite{Houck2012, Hartmann2016}, trapped ions~\cite{Monroe2021sim, Blatt2012}, neural atoms \cite{Browaeys2020, Gross2017}, and photons~\cite{Aspuru-Guzik2012, Hartmann2016}. These systems have been used to study non-trivial quantum effects~\cite{Daley2022,Bloch2005,Bloch2008,Bloch2012,Blatt2012,Oliver2020}, e.g. reproducing phase transitions in quantum many-body systems~\cite{Bloch2005,Bloch2008, Blatt2012,Lukin2017,Lukin20192}. While the analog simulators have arguably reached the practical quantum advantage threshold, the scope of their applications is likely to remain limited to a class of models that can be simulated and the level of precision in quantitative predictions~\cite{Daley2022}. Another approach is to use digital quantum devices~\cite{Lloyd1996} capable of universal quantum computation and in principle not limited in the type of systems they can describe \cite{Martinez2016, Lanyon2011}. Digital quantum simulation can address various physical~\cite{Bassman2021, Bravyi2020sim, Nori2014, Martinis2015} and chemical~\cite{Cao2019, Aspuru-Guzik2020, McArdle2020} problems intractable for classical computing. However, reaching sufficient precision in quantitative predictions calls for significant improvements in the quantum hardware, and likely requires fault-tolerance~\cite{Troyer2021}.
	
Modern quantum computing devices are designed to perform reversible operations and natively support only unitary gates~\cite{Bennett2000}. Simulation of standard Hermitian Hamiltonians fits well within this framework~\cite{Nori2009,Cirac2012,Nori2014}, yet modeling the behavior of non-conservative quantum systems is equally valuable. Understanding Markovian and non-Markovian dynamics of open quantum systems \cite{Rivas2011, Lidar2019, Ashida2020, Luchnikov2022} is important to describe a range of physical phenomena, such as decoherence~\cite{Schlosshauer2019}, thermalization~\cite{DAlessio2015, Nandkishore2015, Reichental2017, Znidaric2009,Abanin2019}, noise characterization~\cite{Harper2020, Youssry2020, Georgopoulos2021}, and others as well as for realizing quantum control protocols ~\cite{James2021, Dong2009,Luchnikov2023}. It is in fact possible to simulate non-unitary dynamics using a reversible quantum computer, and numerous techniques have been developed to this end, including methods based on linear combination of unitaries~\cite{Wei2016, Schlimgen2021, Zheng2021} or dilation~\cite{Hu2020, Head-Marsden2021, Hu2021, Sweke2016}. Effectively, the Hilbert space can be split into two parts -- one part encoding the system of interest, and the other the environment. An interaction between the system and the environment is then simulated by a properly engineered unitary evolution of the total system. 
 
A remarkable special case of non-unitary dynamics arises in parity-time or $\P\T$-symmetric quantum systems~\cite{Bender1997, Bender2007a}. A $\P\T$-symmetric system is described by a Hamiltonian that is non-Hermitian, yet can feature a real energy spectrum. Such systems have properties intermediate between closed and open \cite{Bender2007}, and allow one to realize tunable transitions between the two. Many aspects of $\P\T$-symmetric systems, including those related to information flow~\cite{Kawabata2017, Ju2019, Lee2014}, quantum state discrimination~\cite{Bender2010, Yoo2011}, breaking of entanglement monotonicity~\cite{Chen2014}, have no counterparts in unitary dynamics. However, their distinguishing feature is the phase transition, associated with the breaking of the $\P\T$-symmetry, which is accompanied by a plethora of peculiar physical and mathematical effects. The spectrum of the $\P\T$-symmetric Hamiltonian is real in the unbroken phase, but complex in the broken phase. At the crossover, known as the exceptional point, the complex-conjugated eigenvalues become equal, while the corresponding eigenvectors coalesce~\cite{Heiss2004}. Near the exceptional point, the energy spectrum of the system shows increased response to perturbations, a property that has been proposed as a basis for sensing and signal-processing~\cite{Wiersig2016, Liu2016, El-Ganainy2018}.

Physically, systems with $\P\T$ symmetry can be realized by including suitably balanced gains and losses \cite{El-Ganainy2018}. A natural way to engineer a $\P\T$-symmetric system then is to introduce carefully tuned dissipative couplings. This approach has been demonstrated with a variety of experimental setups including photonics \cite{Ruter2010, Klauck2019, Xiao2021, Klauck2021}, nuclear spins \cite{Zheng2013}, superconducting circuits \cite{Quijandria2018, Naghiloo2019}, and cold atoms \cite{Li2019}. A digital simulation has the potential to be more robust and scalable, as the total system remains unitary and well-controlled. Digital simulation of quantum $\P\T$-symmetry breaking has been demonstrated with the use of nitrogen-vacancy centers in diamonds~\cite{Wu2019} and superconducting qubits~\cite{Dogra2021}.

Our work reports a proof-of-principle experiment simulating the simplest non-trivial $\P\T$-symmetric two-level system using digital unitary evolution of {\it a single three-level quantum system} -- a qutrit. Two of the three qutrit levels correspond to the subspace of the non-Hermitian qubit, while the single remaining level proves sufficient to engineer the effective $\P\T$-symmetric dynamics. As a result, in our setup, the degrees of freedom corresponding to the qubit and the environment are not spatially separated, and the simulation protocol only relies on local single-qutrit gates. This is in contrast to the approach of Refs.~\cite{Wu2019, Dogra2021}, where the environment is represented using ancilla qubits, and interactions are affected by multi-qubit gates. 
 
Generally, multi-level systems (\textit{qudits}) have distinct advantages over qubit systems in the context of quantum information processing~\cite{Farhi1998,Kessel2002,Nielsen2002,Berry2002,Klimov2003,Bagan2003,Vlasov2003,Clark2004,Ralph2007,White2008,Ionicioiu2009,Ivanov2012, Kiktenko2015,Kiktenko2015-2,Song2016,Kiktenko2016,Bocharov2017,Gokhale2019,Pan2019,Low2020,Martinis2009,White2009,Straupe2010,Wallraff2012,Mischuck2012,Martinis2014,Ustinov2015,Morandotti2017,Balestro2017,Low2020,Sawant2020,Senko2020,Pavlidis2021,Rambow2021,Sanders2020, OBrien2022, Gao2022, Cervera-Lierta2021, Galda2021}. In particular, decompositions of multi-qubit gates making use of auxiliary qudit levels~\cite{Ralph2007,White2009,Ionicioiu2009,Wallraff2012}, is an active area of research~\cite{Gokhale2019, Kiktenko2020, Nikolaeva2022}. Significant advantages in quantum simulation, such as a reduction in circuit depth and gate errors in comparison to a traditional qubit-based approach, are also expected (see, e.g., recent proposals presented in Refs.~\cite{Zoller2022, cao2023encoding} and reviews~\cite{Sanders2020,Kiktenko2023}). 

Various physical platforms supporting qudit-based computing are being developed~\cite{Hill2021,roy2022realization,Fischer2023,Ringbauer2022,Kolachevsky2022,OBrien2022}. In particular, superconducting circuits~\cite{Hill2021,roy2022realization,Fischer2023,Nguyen2023} and trapped-ion-based devices~\cite{Ringbauer2022,Kolachevsky2022} have demonstrated promising capabilities. In our work, we use both these leading platforms, operating in the qutrit regime in order to demonstrate a parity–time symmetry breaking in a two-level system.
    
The paper is organized as follows. In Sec.~\ref{sec:PT}, we introduce a two-level $\P\T$-symmetric system, describe its basic properties, and explain how to simulate it digitally using the dilation technique. Secs.~\ref{sec:setup} and \ref{sec:setup-transmon} describe the experimental setup and results for the trapped ion and superconducting platforms, respectively. Sec.~\ref{sec:results} contains discussion and outlook. 

\section{$\P\T$-symmetric systems and simulation}\label{sec:PT}

A $\P\T$-symmetric system is governed by a non-Hermitian Hamiltonian $H$, which is invariant with respect to the combined parity $\P$ and time-reversal $\T$ transformations, $[H, \P\T]=0$.  The characteristic polynomial of a $\P\T$-symmetric Hamiltonian is always real, and hence the eigenvalues are either all real or come in complex-conjugate pairs. In the former case, the system is said to be in the $\P\T$-unbroken phase.  The regime with complex eigenvalues corresponds to the $\P\T$-broken phase and typically arises as the gain and loss terms become sufficiently strong, so that the non-unitary aspects of the dynamics dominate the internal interactions \cite{bender2018pt}.
	
$\P\T$-symmetric systems feature many unusual properties such as complex spectrum, exceptional points and coalescence of eigenvectors, non-conservation of the trance distance between quantum states, and breaking entanglement monotonicity. In this work, we focus on probing the $\P\T$-symmetry breaking phase transition, and the associated qualitative change in the dynamics.
	
\subsection{Two-level $\P\T$-symmetric system}
 The simplest $\P\T$-symmetric system has two levels (qubit) and its time evolution is generated by an effective non-hermitian Hamiltonian (written here with $\hbar$ set to 1)
	\begin{align}
		H=\sigma_x+ir \sigma_z  =\begin{pmatrix}ir & 1 \\ 1 & -ir\end{pmatrix} \ . \label{def H}
	\end{align}
 We henceforth refer to $H$ simply as the $\P\T$-symmetric Hamiltonian.
 The real parameter $r$ quantifies the strength of the gain and loss (diagonal) terms compared to the inter-level interactions. The parity operator is $\P=\sigma_x$, and the time-reversal operator acts by complex conjugation $\T(\cdot)=(\cdot)^*$. The Hamiltonian \eqref{def H} is $\P\T$ symmetric for any real value of $r$, i.e. $[H, \P\T]=0$.
	
The eigenvalues of $H$ are 
	%\begin{align}
		$h_{\pm}=\pm h, h=\sqrt{1-r^2} \ . \label{def h}
	$
 %\end{align}
For $r<1$ the eigenvalues are real and $\P\T$-symmetry is unbroken, while $r>1$ leads to purely imaginary values of $h$ and hence breaks $\P\T$-symmetry. The value $r=1$ corresponds to the exceptional point.
	
Similarly to the unitary case, in the $\P\T$-unbroken phase eigenvectors of the system $|\psi_{\pm}\rangle$ acquire complex phases during the time evolution, and level populations manifest Rabi-like oscillations. {As the phase transition point $r=1$ is approached from within the $\P\T$-unbroken phase, the period of oscillations $T=\frac{2\pi}{\sqrt{1-r^2}}$ grows and diverges at $r=1$. After that, the $\P\T$-broken regime with $r>1$ exhibits an exponential relaxation  to the ground state with time $\tau=\frac1{\sqrt{r^2-1}}$, without the oscillatory behavior \cite{Kawabata_2017}. Our main goal in this work is to probe this expected transition experimentally.
	
\subsection{Embedding non-Hermitian evolution into a unitary operator}

The evolution operator $V(t)=e^{-iHt}$ of a non-Hermitian system 
is not unitary, and hence can not be implemented directly with a reversible quantum computer. 
However, it can be embedded in a unitary gate acting on a larger system
\begin{align}
	U(t) = \begin{pmatrix}\lambda^{-1}V(t) & B \\ C & D \end{pmatrix} \ . \label{def U}
\end{align}
	
Here $\lambda$ is a scalar factor and $B, C, D$ are arbitrary matrix entries compatible with the unitarity of $U(t)$. Such embeddings arise in many settings. Stinespring dilation of CPTP maps is an example~\cite{Stinespring1955}. In the context of quantum algorithms based on transformations of singular values, they are known as block encodings~\cite{Gilyen2018, Martyn2021}. 
An arbitrary operator $V(t)$ can be represented in the form of Eq.~\eqref{def U}, as long as its operator norm satisfies $||V(t)||\le 1$ (for details see Appendix~\ref{app block}). Operators with larger norms can be embedded if rescaled appropriately $V(t)\to V(t)/\lambda$, as we indicated in Eq.~\eqref{def U}.
	
To apply the evolution operator $V(t)$ to an arbitrary initial state $|\psi\rangle$, one embeds $|\psi\rangle$ into the larger space and applies $U(t)$ to the result
	\begin{align}
		U(t)\begin{pmatrix}|\psi\rangle \\0 \end{pmatrix} = \begin{pmatrix}\lambda^{-1}V(t)|\psi\rangle \\ C|\psi\rangle \end{pmatrix} \ .
	\end{align}
	
To probe the structure of the embedded state $V(t)|\psi\rangle$, the measurements are performed on the full resulting state, and the outcomes lying in the correct subspace are \textit{post-selected}.
	
The success probability of the post-selection is equal to $\lambda^{-2} \langle \psi|V(t)^\dagger V(t)|\psi\rangle$. Hence, it decreases as $\lambda$ grows. From this point of view, it is optimal to choose the minimal $\lambda$ compatible with the restriction $\lambda^{-1}||V(t)||\le 1$, which is solved by $\lambda(t)=\sigma_{\rm max}(t)$, with $\sigma_{\rm max}(t)$ being the largest singular value of $V(t)$. We note that rescaling the evolution operator by the scalar factor $\lambda(t)$ is equivalent to shifting the Hamiltonian by a time-dependent constant
\begin{equation}
    H\to H+i\frac{\log\lambda(t)}{t} \ .
\end{equation}
Such a shift does not alter the physical dynamics in the subspace of interest, it only affects the success probability of the post-selection. The post-selection procedure remains unchanged and leads to identical results for any admissible choice of $\lambda(t)$.

\subsection{Simulating two-level $\P\T$-symmetric system with a unitary qutrit}\label{sec:PT-qutrit}
 %%%% FIGURE 1 %%%%
\begin{figure}
\includegraphics[width=0.98\linewidth]{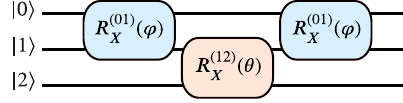}
\caption{Decomposition of the $\P\T$-symmetric qubit dynamics into a sequence of single-qutrit gates.}
\label{fig gates}
\end{figure}

%%%% FIGURE 2 %%%%
\begin{figure*}
  \includegraphics[width=0.98\linewidth]{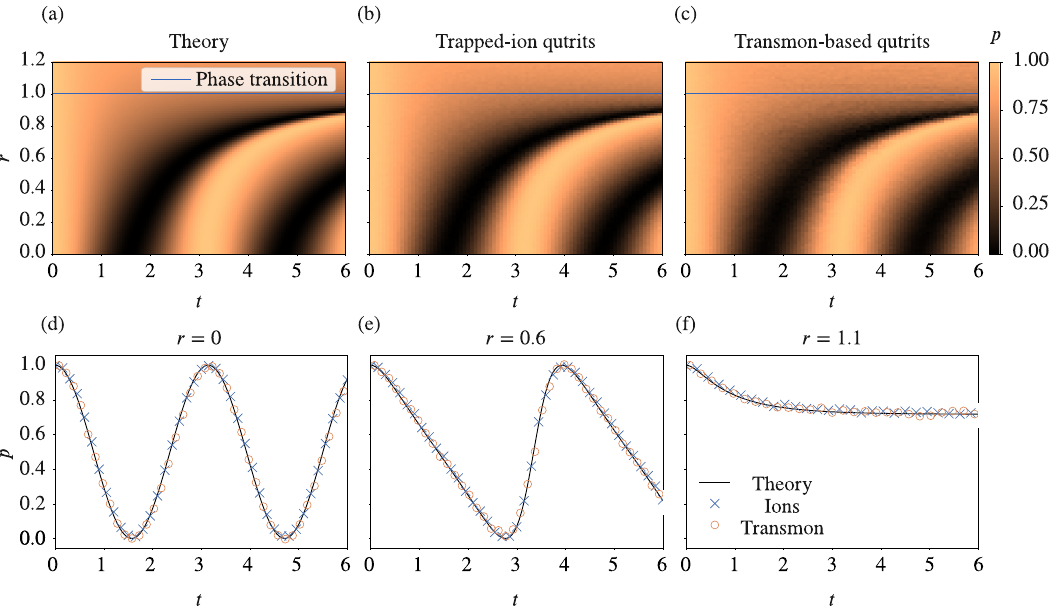}
 % \newline
 \caption{Dynamics ground state population for $\P\T$-symmetric two-level system \eqref{def H} for a range of parameters $0 \leq r \leq 1.2$. The region $r<1$ corresponds to the $\P\T$-unbroken phase, $r>1$ to the $\P\T$-broken phase, and $r=1$ (blue line) to the phase transition (exceptional point). For each point $(r, t)$ rotation angles $(\varphi, \theta)$ are defined according to Eq.~(\ref{eq: angles_varphi_theta}).
(a) Theory.
(b) Experimental results obtained on the trapped-ion platform. Each data point is an average of 8000 experimental runs. The results are SPAM-corrected. 
(c) Experimental results obtained with the transmon-based qutrit. Each data point is an average of 8192 experimental samples. (d) For $r=0$ the evolution is unitary and population dynamics manifests Rabi-like oscillations. (e) Below the exceptional point ($r=0.6$ at the figure) dynamics is non-unitary, but $\P\T$-symmetry unbroken, and level populations are periodic in time. (f) Above the exceptional point ($r=1.1$ at the figure) the $\P\T$-symmetry is broken and level population relaxes exponentially.}
\label{fig:results}
\end{figure*}

The previous section contains a general discussion of embedding a non-unitary evolution operator into larger unitary dynamics. Here we consider the case where the evolution operator is that of the $\P\T$-symmetric qubit Eq.~\eqref{def H}, while the embedding system is a qutrit. There is an additional subtlety in this case, stemming from the fact that a single auxiliary dimension is not sufficient to simulate a general operator $V(t)$. However, precisely for the case when the scalar factor is chosen to be $\lambda(t)=\sigma_{\rm max}(t)$ the embedding is possible, see Appendix~\ref{app block}. 
	
	The evolution operator $V(t)$ for the $\P\T$-symmetric qubit \eqref{def H} can be written as 
    %\begin{equation}
        $V(t)=\cos(ht)-i \frac{\sin(ht)}{h} H,$
    %\end{equation}
    and its singular values read
	\begin{align}
	\sigma_{\pm}(t)=\frac1{|h|}\left(\sqrt{|1-r^2\cos^2(ht)|}\pm |r\sin(ht)|\right) \label{singular values} \ ,
	\end{align}
    so that $\sigma_{\rm max}(t)=\sigma_+(t)$.

The unitary circuit, which corresponds to the target embedding, can be written as a sequence of three elementary qutrit gates (see Fig.~\ref{fig gates}):
\begin{align}\label{UtoXXX}
	U(t)=R_X^{(01)}(\varphi)R_X^{(12)}(\theta)R_X^{(01)}(\varphi) \ , 
\end{align} 
    which are defined by
	\begin{align}
		&R_X^{(01)}(\varphi) = \begin{pmatrix} 
		\cos \frac{\varphi}{2} & -i\sin\frac{\varphi}{2} & 0 \\ 
		-i\sin \frac{\varphi}{2} & \cos\frac{\varphi}{2}  &0 \\
		0 & 0 & 1
		\end{pmatrix} \ , \\
		&R_X^{(12)}(\theta) = \begin{pmatrix} 
		1 & 0 & 0 \\
		0 & \cos \frac{\theta}{2} & -i\sin\frac{\theta}{2}  \\ 
		0 & -i\sin \frac{\theta}{2} & \cos\frac{\theta}{2}
		\end{pmatrix} \ .
		\label{eq:Rx_Ry}
	\end{align}

    %We detail the construction of this circuit in terms of native qutrit gates for both physical platforms below in Secs.~\ref{sec:setup} and \ref{sec:setup-transmon}.
The rotation angles $(\varphi,\theta)$ in Eq.~\eqref{UtoXXX} are functions of the coupling strength $r$ and the evolution time $t$
\begin{align}
	\varphi(r,t)=\arctan\frac{\tan(h t)}{h}, \quad \theta(r,t) = -2\arccos\frac{\sigma_-}{\sigma_+} \ .
\label{eq: angles_varphi_theta}
\end{align}

The return probability $|\langle{0|U(t)|0\rangle}|^2$ computed analytically displays the hallmark phase transition pattern of the $\P\T$-symmetry breaking, Fig.~\ref{fig:results}. 
%Below we describe how to probe this effect experimentally using two experimental platforms: trapped-ion and superconducting transmon-based qutrits. 

\section{Demonstration with trapped-ion qutrits} \label{sec:setup}
    
Here we report the simulation using a trapped ion quantum processor, which is an upgraded version of the recently presented setup (see Refs.~\cite{Kolachevsky2022,Zalivako2023}). It is a chain of ten $^{171}$Yb$^{+}$ ions inside a linear Paul trap. Qudits are encoded in Zeeman sublevels of $^2S_{1/2}(F=0)$ and $^2D_{3/2}(F=2)$, with the qudit dimension up to $d=6$. In this work we employ only three states of each qudit, which we further refer as $|0\rangle=\,^2S_{1/2}(F=0, m_F=0)$, $|1\rangle=\,^2D_{3/2}(F=2,m_F=0)$ and $|2\rangle=\,^2D_{3/2}(F=2,m_F=1)$. More details on the experimental setup are given in the ~Appendix~\ref{app:ions}. Information about initialization, quantum gates, and readout procedures are also given there.

Native single-qudit operations supported by our processor~\cite{Kolachevsky2022} are $R^{(0j)}(\varphi, \theta)$ and \textit{virtual} $R_Z^{(0j)}(\theta)$  gates with $j = 1,2$. Their matrix representations are given in  Appendix~\ref{app:ions}. The virtual $R_Z$ gates are not used in the current experiment, and will not be discussed in detail here. $R_X$-rotations featuring in decomposition \eqref{UtoXXX} can be transpiled to the native gates using relations $R_X^{(0i)}(\theta)=R^{(0i)}(0, \theta)$ and 
\begin{align}
    R_X^{(ij)}(\theta) = 
    R_Y^{(0i)}(\pi)
    R_X^{(0j)}(\theta)
    R_Y^{(0i)}(-\pi)\ ,
\end{align}
where $R_Y^{(0i)}(\theta)=R^{(0i)}(\pi/2, \theta)$. The result of the transpilation is given in Fig.~\ref{fig:transpiled-circuit}.

%%%% FIGURE 3 %%%%
\begin{figure}[h]
\includegraphics[width=0.98\linewidth]{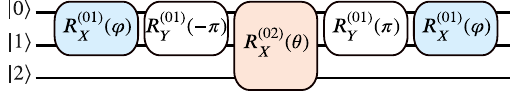}
\caption{Simulation circuit \eqref{UtoXXX} transpiled to the single-qutrit gates native to the trapped ion processor.}
\label{fig:transpiled-circuit}
\end{figure}
%%%% FIGURE 4 %%%%
\begin{figure*}
\includegraphics[width=0.98\linewidth]{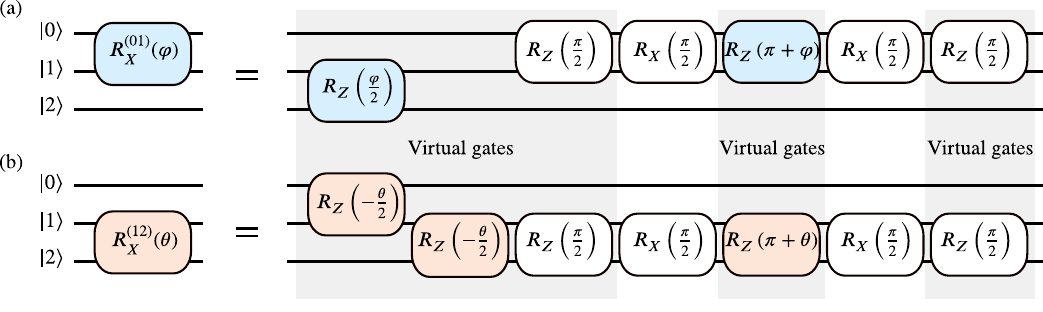}     
\caption{Gates (a) $R_X^{(01)}(\varphi)$ and (b) $R_X^{(12)}(\theta)$ transpiled to the native single-qutrit transmon operations. Virtual $R_Z$ gates are highlighted with a gray background. Only the colored gates are subject to the parameter sweeps in the simulation protocol of Fig.~\ref{fig gates}.}
\label{fig:transmon_exp_gates}
\end{figure*}
    
As mentioned, ten ion qudits are available in our setup, and the addressing laser system enables us to control each ion individually. Since the experiment only involves single-qudit operations, we chose to increase the sampling rate by performing the parallel computation on 5 out of 10 ions. We chose to use only half of the ions (so that no active ions are nearest neighbors), to reduce the cross-talk effects.

Experimental results obtained with the trapped ion processor are shown in Fig.~\ref{fig:results}(b).
Each pair of parameters $(r, t)$ defines rotations angles $(\varphi,\theta)$ for the transpiled circuit Fig.~\ref{fig:transpiled-circuit} according to Eq.~(\ref{eq: angles_varphi_theta}), and 8000 samples are aggregated and averaged to compute level populations for each datapoint. Results are post-selected on lying in the qubit subspace $|0\rangle, |1\rangle$, yielding probabilities $p_0(r, t)$ and $p_1(r, t)$. The return probability of the non-hermitian qubit is then computed as
\begin{align}
    \frac{p_0 (r, t)}{p_0 (r, t) + p_1 (r, t)} \ ,
\end{align}
and is the final quantity reported in Fig.~\ref{fig:results}(b).

\section{Demonstration with a transmon-based qutrit} \label{sec:setup-transmon}

The transmon-based qutrit is used to access the three-level system with a superconducting platform. The transmon is a widely used qubit consisting of a Josephson junction shunted with large capacitance~\cite{koch2007charge}. It has an energy spectrum of a weakly anharmonic quantum oscillator, which allows using it as a qutrit. 
For details on the device, initialization procedure, gate implementation, and readout see Appendix~\ref{transmon setup}.
The native gate set for our superconducting qutrit consists of $R_X^{(01)}(\varphi)$, $R_X^{(12)}(\varphi)$,  $R_Z^{(01)}(\varphi)$, $R_Z^{(12)}(\varphi)$ rotations, their matrix representations also given in Appendix~\ref{transmon setup}.

While it is possible to implement gates $R_X^{(01)}(\varphi)$ and $R_X^{(12)}(\varphi)$ operations with an arbitrary angle $\varphi$, each value of $\varphi$ requires preliminary measurement-intense calibration. In turn, the gates $R_Z^{(01)}(\varphi)$ and $R_Z^{(12)}(\varphi)$ can be implemented \textit{virtually} for any $\varphi$ with zero duration and perfect fidelity \cite{mckay2017efficient}. In terms of a total calibration time reduction, it is more efficient to transpile the gate sequences using $R_X$ gates with fixed angles and arbitrary $R_Z$ rotations. In this work, we calibrate and use $R_X^{(01)}(\pi/2)$ and $R_X^{(12)}(\pi/2)$, which form a universal single-qutrit gate set when supplemented with the virtual $R_Z$ rotations.

To transpile Eq.~\eqref{UtoXXX} into this gate set, we use a relation
\begin{align}
    & R_X^{(j~ j+1)}(\varphi)=\left(e^{-i\varphi / 2}\right)^{(j~ j+1)} H^{(j~ j+1)} R_Z^{(j~ j+1)}(\varphi) H^{(j~ j+1)}
    \label{eq:transmon_qutrit_RX_transpilation} \ ,
    \end{align}
where $j \in  \{0, 1\}$, and $H^{(j~ j+1)}$ denotes an operation similar to a Hadamard gate: 
\begin{align}
    H^{(j~ j+1)}=R_Z^{(j~ j+1)}\left(\frac{\pi}{2}\right) R_X^{(j~ j+1)}\left(\frac{\pi}{2}\right) R_Z^{(j~ j+1)}\left(\frac{\pi}{2}\right).
\end{align}
    
We note that in Eq.~(\ref{eq:transmon_qutrit_RX_transpilation}) the global phase $\left(e^{i\lambda}\right)^{(j~ j+1)}$ of a two-level subsystem phase cannot be left out, but can be reproduced by a combination of two-level $R_Z$ rotations
\begin{align}
 & \left(e^{i\lambda}\right)^{(01)} = R_Z^{(01)}\left(0\right)R_Z^{(12)}\left(-\lambda\right) \ ,\\
& \left(e^{i\lambda}\right)^{(12)} = R_Z^{(12)}\left(\lambda \right)R_Z^{(01)}\left(\lambda\right) \ .
\end{align}

Fig.~\ref{fig:transmon_exp_gates} depicts the transpilation of the gates $R_X^{(0 1)}(\varphi)$ and $R_X^{(1 2)}(\theta)$, featuring in the simulation circuit Fig.~\ref{fig gates}.
    
Experimental results obtained with a superconducting platform agree with the theoretical predictions and are reported in Fig.~\ref{fig:results}(c). The parametrization of rotation angles $ \varphi (r, t)$ and $ \theta (r, t)$ is used in the transpiled gates in Fig.~\ref{fig:transmon_exp_gates} and post-selection of the outcome probabilities are similar to the ion-based experiment.

\section{Discussion}   

Here we give a summary of the experimental results demonstrating the $\P\T$-symmetry breaking on both experimental platforms Fig.~\ref{fig:results}, and discuss some of their specific features. For both setups experimental data are very close to the theoretical model. In the trapped-ion case some statistical noise is present, consistent with 8000 samples per point averaging.

For the transmon-based device, each reported observation value is an average of 8192 experimental sequences, preparing the state populations of a superconducting qutrit. It should be noted that a phase increment value is discretized in our waveform generator, therefore one can notice a slight ripple behavior in  Fig.~\ref{fig:results}c. We also note that, though the transpiled circuit in~Fig.~\ref{fig:transmon_exp_gates} looks much longer than the original one, it mostly consists of virtual zero-duration $R_Z$ rotations (highlighted with gray boxes). Hence, the total circuit duration is comparable to the original sequence.

We would like to point out the essential differences between our experiments and the previous works, where the parity-time symmetry transition has been observed. Similarly to our setup, Ref.~\cite{Naghiloo2019} used three levels of a transmon to probe the phase transition. However, their simulation is analog, relies on engineered dissipative couplings and controlled relaxation of the excited states. In particular, this technique has the drawback of post-selection sucess rate decaying exponentially with the simulation time. Our simulation is digital, allowing more control, and the post-selection success rate does not decay with time. Similarly to our work, Ref.~\cite{Dogra2021} relies on the fully digital simulation, but uses auxiliary qubits to engineer non-hermiticity. 
As we have argued in Introduction, multi-level systems can provide distinct advantages in storing and and processing of quantum information. Illustrating this potential, our work uses a single qutrit -- the minimal possible setup to probe this phase transition digitally, and the techniques developed apply broadly.

\section{Conclusions} \label{sec:results}

We have introduced a theoretical protocol for the simulation of a two-level $\P\T$-symmetric system using the digital evolution of a unitary qutrit. The simulation is based on the dilation technique, i.e. embedding of the non-unitary evolution operator into the unitary dynamics of a larger system. A single additional level existing in a qutrit proved to be sufficient for our application.
The protocol has been implemented in two independent experimental setups -- trapped ions and a superconducting transmon, and conclusively demonstrated the predicted change in dynamics across the $\P\T$-symmetry breaking phase transition, from oscillatory behavior ($\P\T$-symmetric regime) to the exponential relaxation ($\P\T$-broken regime). 
Both experimental platforms have demonstrated excellent agreement with each other and with the theory. 
Our results point to the significant potential of both, trapped ions and superconductors, in simulating the physics of open systems. 

\section*{Acknowledgements}
We acknowledge A. Ustinov for fruitful discussions and comments on the manuscript. 
The work was supported by Rosatom in the framework of the Russian Roadmap for Quantum computing (Contract No. 868-1.3-15/15-2021 dated October 5, 2021 and Contract No. 151/21-503 dated December 21, 2021). 
The work of N.A.N, A.S.N., and A.K.F. was supported by the Federal Academic Leadership Program Priority 2030, Strategic Project Quantum Internet (Project No. K1-2022-027 at MISIS). 
N.A.N. thanks the support of the Russian Science Foundation Grant No. 23-71-01095 (theoretical modeling of quantum circuits). 
E.O.K. thanks for the support of the Russian Science Foundation Grant No. 19-71-10091 (study of embedding qubits’ non-Hermitian dynamics into qutrits). 
The transmon-based device was supported by Rosatom and fabricated using the equipment of MIPT Shared Facilities Center (Contract No. 868/221-D dated October 24, 2022).

N.A.N., A.G., and A.K.F. proposed an idea for the project.
N.A.N., A.S.N., E.O.K., and A.K.F. worked on the theoretical analysis.
I.V.Z., A.S.B., K.P.G., and A.E.K. performed experimental work using the trapped-ion setup with the conceptual contribution from N.N.K., K.Yu.K., and I.A.S. N.V.S. contributed to the development of the single-shot ion qudit readout and data analysis.
A.S.K. developed a way of experimental realization of a transmon-based qutrit.
A.S.K., I.A.S., A.V.K., E.Yu.E., and N.N.A., performed experiments on a transmon-based qutrit.
A.S.K. and E.Yu.E. designed a sample of superconducting transmon, while
D.K., V.L., and A.N.B. fabricated it. N.M. supervised the project on the superconducting group.
N.N.K., K.Yu.K., and I.A.S. supervised the project on the trapped-ion group.
N.A.N., A.S.N., I.V.Z., A.S.K., I.A.S., N.M., and A.K.F wrote the manuscript with the contribution of other coauthors. 
A.K.F. supervised the project.

\appendix
	
\section{Block encoding} \label{app block}
 \subsection{General}

    To make the technical aspects of our work self-contained, here we present a simple approach to block encodings of non-unitary operators. Let $A$ be an $n\times n$ operator that we wish to embed into an $(n+m)\times(n+m)$-dimensional unitary $U$, with the following block structure
	\begin{align}
	U=\begin{pmatrix} A_{n\times n} & B_{n\times m}\\ C_{m\times n} & D_{m\times m}\end{pmatrix} \ .
	\end{align}
    In our applications, $A$ is the evolution operator $A=e^{-iHt}$ of some non-hermitian Hamiltonian. We would like to stress that $\P\T$-symmetry is not required at this point, and the technique applies to general non-hermitian Hamiltonians.
    
    We assume to have the full control over the $n+m$-dimensional system, so that the only constraint on $A$ comes from the unitarity of $U$, i.e. $U^\dagger U=\mathbb{1}$, or explicitly
	\begin{align}
	A^\dagger A+C^\dagger C=\mathbb{1},\quad A^\dagger B+C^\dagger D=0 \ ,\\
	B^\dagger A+D^\dagger C=0,\quad B^\dagger B+D^\dagger D=\mathbb{1} \ . \label{U constraints}
	\end{align}
 
    Assuming that the first of these equations can be solved for $C$, the remaining equations have solutions as well. Indeed, the off-diagonal equations are solved by choosing
	\begin{align}
	B=-(A^\dagger)^{-1}C^\dagger D \ .
	\end{align}
	Note since $A$ is an exponential, $A^{-1}$ exists. Substituting $B$ into the last equation leads to
	\begin{align}
	D^\dagger KD=\mathbb{1},\quad K=CA^{-1}(CA^{-1})^\dagger+\mathbb{1} \ .
	\end{align}
	Because $K$ is Hermitian it can always be diagonalized by a unitary transformation $W^\dagger K W=\operatorname{diag}(k_1,k_2,\dots)$. Since $K$ is positive-definite $k_i>0$. Hence, choosing
	\begin{align}
	D=W \operatorname{diag}\left(\frac{1}{\sqrt{k_1}},\frac{1}{\sqrt{k_2}},\dots\right)
	\end{align}
	fulfills the last equation in Eq.~\eqref{U constraints}.
	
	Thus, the key question is whether the first equation in Eq.~\eqref{U constraints} has a solution. In fact it does, provided 
     \begin{itemize}
     \item[] (i) $\mathbb{1}-A^\dagger A\ge0$ .
     \item[] (ii) $\operatorname{rank}(\mathbb{1}-A^\dagger A)\le m$.
     \end{itemize}
    The first condition here ensures that $C^\dagger C$ is positive semi-definite, and is equivalent to the requirement $||A||\le 1$. The second condition takes into account the fact that $m$ is the maximum rank of $C^\dagger C$, for $m\times n$-dimensional operator $C$. An explicit solution for $C$ can be given e.g. in the basis diagonalizing $A^\dagger A$, but we will not need it.

    \subsection{Relaxing restrictions} \label{sec relax}
	Constraints on the singular values of $A$ might appear to be too restrictive in practice. For example, $||A||=||e^{-iHt}||\le 1$ generally would not hold for evolution operators in non-Hermitian systems. A simple work-around is to instead simulate $H+i\mu$ with some sufficiently large real constant $\mu$. Shifting the Hamiltonian by a constant affects the dynamics of the physical subspace trivially, but permits a block encoding into a unitary matrix.

    Assume that the largest singular value $\sigma_{\rm max}$ of $A$ is known. Then block encoding $A/\sigma_{\rm max}$ is a natural choice. It puts all eigenvalues in the range $[0, 1]$, and at the same time reduces the rank of $\mathbb{1}-A^\dagger A$ by one, allowing to use one less auxiliary dimension for block embedding. The last property is important for this work, since it allows simulating an arbitrary two-dimensional system using a single extra dimension, i.e. a unitary qutrit.

    \subsection{Gate-level implementation}
    An arbitrary single-qutrit gate, i.e. an element $U\in SU(3)$, can be decomposed into a product of three two-level gates 
    \begin{align}
     U=A^{(01)}B^{(12)}C^{(01)}   \ , \label{U to ABC}
    \end{align}
    where $A, B, C \in SU(2)$. For a simple proof, we refer to the appendix of Ref. ~\cite{Rowe1999}. As we now show, the additional symmetries of the $\P\T$-symmetric Hamiltonian \eqref{def H} lead to a very compact form of the decomposition.

We begin by observing that
\begin{align}
e^{-iHt}=\cos(ht)-i\frac{\sin(ht)}{h}H\ ,
\end{align}
and rewrite it as
\begin{align}
e^{-iHt}=\frac{\sqrt{1-r^2\cos^2(ht)}}{h}e^{i\varphi \sigma_x}+r\frac{\sin(ht)}{h}\sigma_z \ .    
\end{align}
Here 
\begin{align}
\varphi = \arctan \frac{\tan(ht)}{h}, \quad h=\sqrt{1-r^2}\ .
\end{align}
This form naturally leads to the following singular value decomposition
\begin{align}
& e^{-iHt}=R_X(\varphi)\Sigma R_X(\varphi) \label{RSR} \ ,\\
& \Sigma= \frac{1}{h}\left(\sqrt{1-r^2\cos^2ht}+r\sin ht\,\,\sigma_z\right) \ , 
\end{align}
where $R_X(\varphi)=e^{-i\frac12\varphi \sigma_x}$. Hence, the singular values are
\begin{align}
\sigma_{\pm}=\frac1h\left(\sqrt{1-r^2\cos^2ht}\pm r\sin ht\right) \ .
\end{align}
Note that $\sigma_{\pm}\ge0$ for all $r\in\mathbb{R}$, and can alternatively be written as shown in Eq.~\eqref{singular values}.

The renormalized evolution operator to be embedded into a qutrit unitary is $e^{-iHt}/\sigma_{\rm max}$. Its singular values are 1 and $\sigma$
\begin{align}
\sigma=\frac{\sigma_{\rm min}}{\sigma_{\rm max}}=\frac{\sqrt{|1-r^2\cos^2ht|}-|r\sin ht|}{\sqrt{|1-r^2\cos^2ht|}+|r\sin ht|} \ .
\end{align}

Decomposition of the form \eqref{U to ABC} can now be derived from the factorization \eqref{RSR} (see Fig.~\ref{fig gates} for the graphical representation)
\begin{align}
U=R_X^{(01)}(\varphi)R_X^{(12)}(\theta)R_X^{(01)}(\varphi) \ ,
\end{align}
with $\theta=-2\arccos\sigma$. The middle factor here reads
\begin{align}
R_X^{(12)}(\theta) = \begin{pmatrix}1 & 0 & 0 \\ 0 & \cos\frac{\theta}{2} & -i\sin\frac{\theta}{2} \\  0 & -i\sin\frac{\theta}{2} & \cos\frac{\theta}{2}\end{pmatrix} = \begin{pmatrix}1 & 0 & 0 \\ 0 & \sigma & * \\ 0 & * & * \end{pmatrix} \ .
\end{align}
The last form emphasizes that the top left block of $R_X^{(12)}(\theta)$ reproduces $\Sigma/\sigma_{\rm max}$, while the unspecified entries $*$ do not affect the resulting block encoding. They can be chosen arbitrarily (subject to the unitarity constraint), and $R_X^{(12)}(\theta)$ provides perhaps the simplest such choice.

\section{Details on trapped-ion-based qutrit} \label{app:ions}
    \subsection{Initialization}
    Basic scheme of the trapped-ion setup is given in the Fig.~\ref{fig:ion_setups}. At the beginning of each experimental run, ions are first Doppler cooled with a combination of 369.5~nm phase-modulated at 14.7~GHz laser and a 935.2~nm phase-modulated at 3.08~GHz laser~\cite{zalivako2019improved} (Fig.~\ref{fig:app-ion-levels}). After that each qudit is initialized in the $|0\rangle$ state by the optical pumping with the same lasers (369.5~nm laser phase modulation frequency is changed to 2.1~GHz for that). Usually, after this step ions radial modes are sideband cooled to the motional ground state~\cite{Monroe1995}, which is required for two-qudit operations, but in this experiment this step is omitted as only single-qudit operations are necessary.

    %%%% FIGURE 5-1 %%%%
    \begin{figure}
        \includegraphics[width=0.98\linewidth]{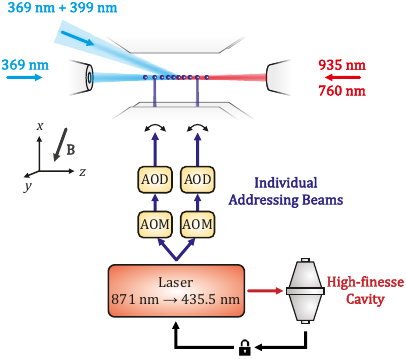}
         \caption{Simplified scheme of the trapped-ion setup. Ions are stored in a linear Paul trap. Beams at 369~nm, 935~nm and 760~nm ensure ions cooling, initialization, readout and repumping. Quantum operations are performed with two tightly focused laser beams at 435~nm, which can be scanned with acousto-optical deflectors (AOD) along the ion chain. Acousto-optical modulators (AOM) control their amplitude, phase and frequency. The addressing laser frequency is stabilized with respect to a high-finesse optical cavity.}
          \label{fig:ion_setups}
    \end{figure}
    
    %%%% FIGURE 5 %%%%
    \begin{figure}
        \includegraphics[width=0.98\linewidth]{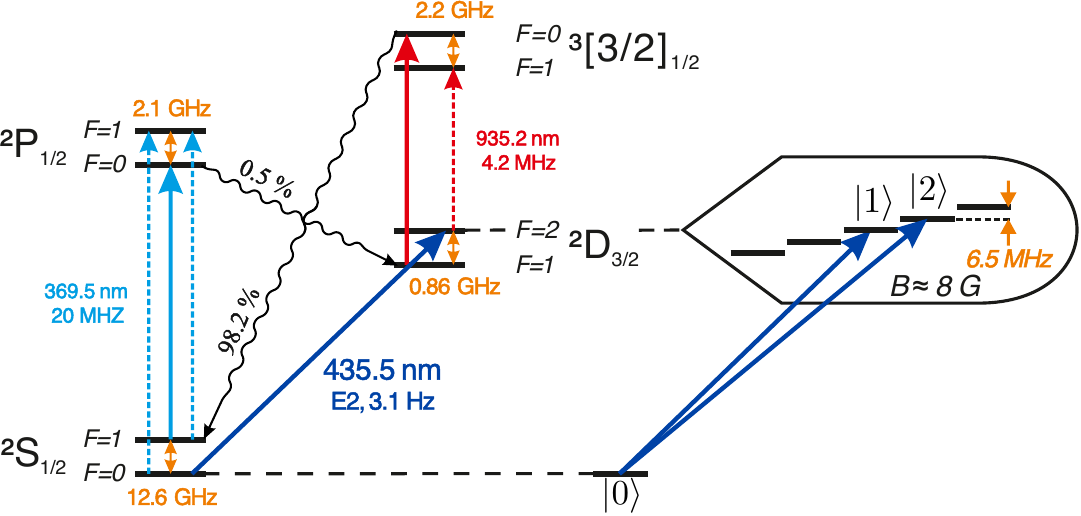}
         \caption{Level scheme of the $^{171}$Yb$^{+}$ ion. Solid lines show laser fields. Dashed lines show laser fields obtained by lasers phase modulation.}
          \label{fig:app-ion-levels}
    \end{figure}
    
    \subsection{Native gates}
    Native gates $R^{(0j)}(\varphi, \theta)$ used in this paper are given by the following matrices:
    \begin{align} 
		&R^{(01)}(\varphi, \theta) = \begin{pmatrix} 
		\cos \frac{\theta}{2} & -ie^{-i\varphi}\sin\frac{\theta}{2} & 0 \\ 
		-ie^{i\varphi}\sin \frac{\theta}{2} & \cos\frac{\theta}{2}  &0 \\
		0 & 0 & 1
		\end{pmatrix} \ , \\
		&R^{(02)}(\varphi, \theta) = \begin{pmatrix} 
		\cos \frac{\theta}{2} & 0 & -ie^{-i\varphi}\sin\frac{\theta}{2} \\ 
		0 & 1 & 0  \\ 
		-ie^{i\varphi}\sin \frac{\theta}{2} & 0  &\cos\frac{\theta}{2} \\
		\end{pmatrix} \ ,
		\label{eq:R_phi_theta}
	\end{align}
 
 The gates $R^{(0j)}(\varphi, \theta)$ are implemented by applying a laser pulse at 435.5~nm resonant to the $|0\rangle \to |j\rangle$ transition. Relative phase of the laser emission sets the angle $\varphi$, while the pulse duration determines $\theta$. 
    \subsection{Readout} \label{app:ion-readout}
    %%%% FIGURE 6 %%%%
        \begin{figure*}
         \includegraphics[width=0.9\linewidth]{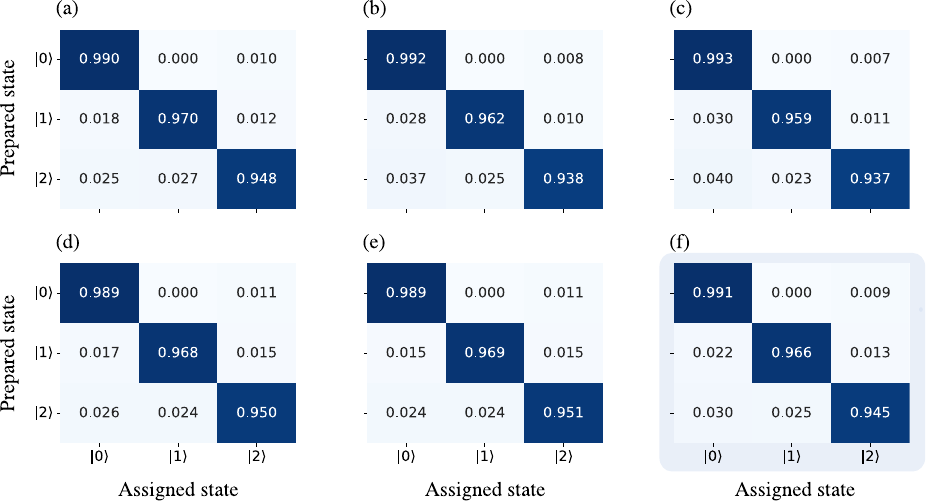}
         \caption{Readout confusion matrices for trapped ion qutrits. (a)-(e) for each of the 5 used ions, (f) average readout confusion matrix for these 5 ions.}
         \label{fig:confusion_matrix_ions}
    \end{figure*}
    After applying required quantum gates a state readout of each ion is performed. The first stage of this procedure is analogous to the optical qubit~\cite{semenin2021optimization}. The ions are illuminated with a 369.5~nm cooling laser phase-modulated at 14.7~GHz and a 935.2~nm non-modulated repumping beam. These fields drive transitions $^2S_{1/2}(F=1) \to \,^2P_{1/2}(F=0)$, $^2S_{1/2}(F=0) \to \,^2P_{1/2}(F=1)$ and $^2D_{3/2}(F=1) \to \,^3[3/2]]_{1/2}(F=0)$ resulting in a strong fluorescence of the ions being in the $|0\rangle$ state in the end of the quantum algorithm. Ions in states $|1\rangle$ and $|2\rangle$ remain dark. Ions fluorescence photons are collected with a high-aperture lens and focused onto an array of multi-mode optical fibers. Other ends of these fibers are connected to the photo-multiplier tubes. By comparing the number of detected photons during the measurement cycle (single-cycle duration is 900~\textmu s) for each ion with a predetermined threshold value, we distinguish state $|0\rangle$ from all others. At the end of the measurement cycle, all population from the state $|0\rangle$ is transferred to the $^2S_{1/2}(F=1)$. After that operation $R^{(01)}(0, \pi)$ is applied to all the ions transferring population from the $|1\rangle$ state to the empty $|0\rangle$ state, and the measurement is repeated. In the second measurement cycle, the ion is dark only if it is in the $|2\rangle$ state at the end of the algorithm. Thus, with these two measurement cycles, we can distinguish all three states of each ion. 

    To calibrate the initialization and readout processes we sequentially prepared all ions in states $|0\rangle$, $|1\rangle$ and $|2\rangle$ and performed the measurement of the register. For each state, $10^4$ measurements are made. In Fig.~\ref{fig:confusion_matrix_ions} we show the confusion matrix for the readout process for all 5 used ions and an average readout fidelity through them.

    The readout error sources include optical pumping from the $|0\rangle$ state to the $^2D_{3/2}(F=2)$ manifold during the transient stage in the beginning of each readout cycle, non-resonant pumping between qudit states, single-qudit gates infidelities and spontaneous decay of the $^2D_{3/2}(F=2)$ states~\cite{semenin2021optimization}. All these error contributions can be rather straightforwardly reduced by further optimization of the system parameters. For instance, it was demonstrated, that a duration of a single readout cycle in multiparticle processors based on $^{171}$Yb$^{+}$ ions can be significantly decreased in comparison to our current results by reducing the amount of the stray light on the detector and increasing a photon collection efficiency~\cite{debnath2016programmable}. This would significantly reduce errors due to spontaneous decay and non-resonant pumping.
    
    This method can also be easily extended to the case where all $d=6$ qudit states are used to encode information.

\section{Details on transmon-based qutrit} \label{transmon setup}
\subsection{Device description}
    %%%% FIGURE 7 %%%%
    \begin{figure}
         \includegraphics[width=0.9\linewidth]{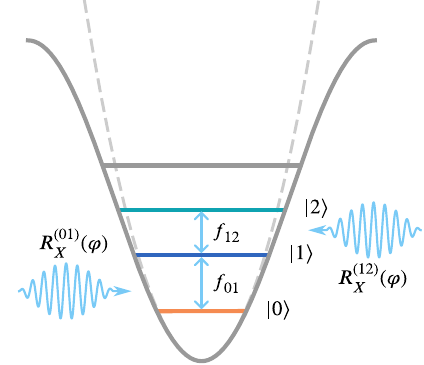}
         \caption{The energy spectrum of a transmon-based qutrit. Computational levels $\ket{0}, \ket{1}$ and $\ket{2}$ of a qutrit system are colored. The allowed transitions are $0-1$ and $1-2$ with corresponding frequencies $f_{01}$ and $f_{12}$. Native gates $R_X^{(01)}(\varphi)$ and $R_X^{(12)}(\varphi)$ are represented as allowed operations of applied microwave pulses.}
         \label{fig:transmon_levels}
    \end{figure}
For the purposes of this research we use a flux-tunable transmon qubit, where the first three energy levels are treated as a qutrit system \cite{Rasmussen2021}, see Fig.~\ref{fig:transmon_levels}. The fabrication process of a such device consist of five main parts: ground plane fabrication, fabrication of $\text{Al}/\text{AlO}_x/\text{Al}$ Josephson junctions, bandages deposition, and air-bridges construction. The fabrication starts with silicon substrate cleaning and aluminum thin film evaporation. The main structures including transmon electrodes and coplanar waveguide transmission lines are patterned using a direct optical lithography. The next step is aimed at the Josephson junction fabrication using standard Dolan bridge technique \cite{dolan1977offset}. In order to have good galvanic contact between the ground plane and the obtained Josephson junctions bandages are deposited through a single-layer organic mask after aluminum oxide etching via an argon milling process. In order to achieve uniform electric potential and avoid parasitic modes, the final fabrication step is devoted to aluminum free-standing air-bridge structures \cite{chen2014fabrication}.

At the operating point (sweet spot), where the energy spectrum is less sensitive to the flux noise, the transition frequencies $f_{01}$ and $f_{12}$ are $6.16$ GHz and $6.04$ GHz respectively. We probe states via the dispersive readout scheme \cite{blais2004cavity} using an individual transmission line resonator of the frequency $7.1$ GHz. Since the qutrit transition frequencies are relatively close to the resonator the transmon can suffer from a spontaneous Purcell decay. Therefore, to protect qutrit, an individual coplanar waveguide filter with wide linewidth is added to the scheme according to the proposal described in~\cite{heinsoo2018rapid}.

We characterize the coherence properties of the qutrit system by measuring the spontaneous decay rates from state $\ket{1}$ ($T_1^{(01)}=10.5~\mu\text{s}$) and state $\ket{2}$ ($T_1^{(12)}=4.8~\mu\text{s}$) and Ramsey oscillations between $\ket{0}~\text{and}~\ket{1}$ ($T_2^{(01)}=6.2 ~\mu\text{s}$).

\subsection{Experimental setup}
%%%% FIGURE 8 %%%%
\begin{figure*}[t]
    \center{\includegraphics[width=0.98\textwidth]{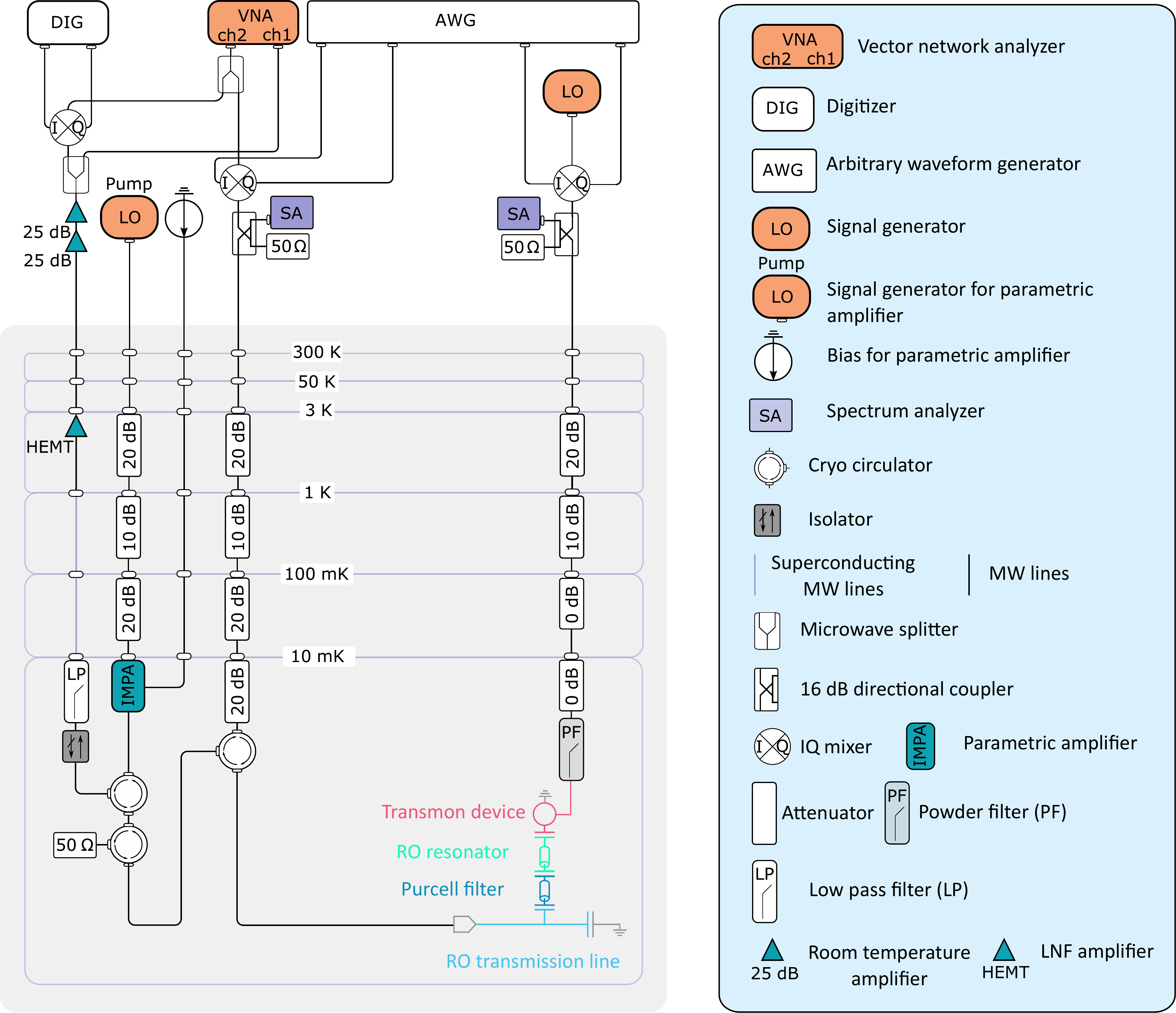}}
    \caption{Experimental setup of the transmon-based experiment.}
    \label{fig:scheme_eq}
\end{figure*}
The presented experiment is performed in the dilution refrigerator with a base temperature of around 10 mK, see Fig.~\ref{fig:scheme_eq}. The whole experimental setup can be divided into two main parts: cryogenic and room temperature. In the dilution refrigerator microwave attenuators are used for thermalisation purposes.
The transmon is coupled to a readout transmission line via a resonator and a Purcell filter, for gate implementations ($XY$ controls) and flux control ($Z$ control) an additional is used. 

Pulse generation for qutrit control is fully performed by an arbitrary waveform generator (AWG) with a local oscillator (LO). The IQ mixers have the ability to combine two signals from AWG, which supply a pulse envelope of a low intermediate frequency component, with a high-frequency signal from LO \cite{Krantz_2019}. One channel from AWG is also used for flux control. The same IQ up- and down-convertion approach is used for qutrit readout. The signal from the transmission line is amplified by an impedance-matching parametric amplifier (IMPA) and then processed with a custom digitizer (DIG) based on FPGA.

The experimental procedures can be divided into three main steps: initialization, single-qutrit gate pulses, and individual readout. Below, we describe each part of the experiment in detail.

\subsubsection{Initialization}

We use the passive reset method and wait for approximately $100 \; \mu$s, allowing the qubit to naturally dissipate into the external environment. The initial state prepared in this way is a good approximation of the ground state $\ket{0}$ in our case, since $h f_{01} \ll k T$, where $T$ is the environmental temperature in the dilution refrigerator, $k$ is the Boltzmann constant and $h$ is the Plank constant. This implies that the residual thermal population can be neglected.

\subsubsection{Single-qutrit gates}

In order to manipulate the qutrit states, we use microwave pulses generated by the standard heterodyne approach \cite{Krantz_2019}. In the lab frame, the Hamiltonian function of the transmon-based qutrit system under the external drive can be written as

\begin{equation}
    \hat{H}_{\text{lab}} = \hbar \sum_{j =1, 2} \left( \omega_j \ket{j} \bra{j} +\lambda_j\Omega(t) (\hat{\sigma}_j^- + \hat{\sigma}_j^+)\right),
    \label{eq: transmon_lab_drive}
\end{equation}
where $\hat{\sigma}_j^- = \ket{j-1} \bra{j}$ and $\hat{\sigma}_j^+ = \ket{j} \bra{j-1}$  are the lowering and the raising operators respectively, $\hbar \omega_j$ is the energy of a state $\ket{j}$. The drive term $\Omega(t)$ with modulation frequency $\omega_d$ is naturally expressed as $\Omega(t) = I(t) \cos \omega_{d}t + Q(t) \sin \omega_{d}t$. Here, we also introduce the weight parameter $\lambda_j$, conditioned by the energy structure of a system. For a transmon, $\lambda_1 = 1, \lambda_2=\sqrt{2}$ due to the charge matrix elements.

In a rotating frame, Eq.~(\ref{eq: transmon_lab_drive}) simplifies to 
\begin{equation}
    \hat{H}_{\text{RWA}} =
    \begin{pmatrix}
    0 & I(t) + iQ (t)& 0\\
    I(t) - iQ(t) & 0 & \sqrt{2}(I(t) + iQ(t))\\
    0 & \sqrt{2}(I(t) - iQ(t)) & 0\\
    \end{pmatrix} \ .
    \label{eq:transmon_rwa_drive}
\end{equation}

Thus we can execute two-level $R_X$ rotations the subspaces spanned by states \{$\ket{0}$, $\ket{1}$\} and \{$\ket{1}$, $\ket{2}$\}, being our first pair of native gates. In the matrix representation these gates are defined as follows~\cite{Morvan_2021, Goss_2022}:
\begin{equation}
R_X^{(01)}(\varphi) = \begin{pmatrix} 
\cos \frac{\varphi}{2} & -i\sin\frac{\varphi}{2} & 0 \\ 
-i\sin \frac{\varphi}{2} & \cos\frac{\varphi}{2}  &0 \\
0 & 0 & 1
\end{pmatrix} \ ,
\end{equation}

\begin{equation}
R_X^{(12)}(\theta) = \begin{pmatrix} 
1 & 0 & 0 \\
0 & \cos \frac{\theta}{2} & -i\sin\frac{\theta}{2}  \\ 
0 & -i\sin \frac{\theta}{2} & \cos\frac{\theta}{2}
\end{pmatrix} \ .
\end{equation}

In-phase $I(t)$ and quarter-phase $Q(t)$  quadratures hold information not only about a pulse shape, but also about the signal modulation phase $\varphi$. It can be shown, that the phase incrementation to the drive modulation gives instantaneous change of rotation axis producing virtual Z-gates. In the matrix representation this pair of our native gates is defined by
\begin{equation}
        R^{(01)}_Z(\varphi)=\begin{pmatrix} 
        1 & 0 & 0 \\
        0 & e^{i \varphi} & 0 \\ 
        0 & 0 & 1
        \end{pmatrix},
        \label{eq:rz_01_transmon}
\end{equation}
\begin{equation}
        R^{(12)}_Z(\varphi)=\begin{pmatrix} 
        1 & 0 & 0 \\
        0 & 1 & 0 \\ 
        0 & 0 & e^{i \varphi}
        \end{pmatrix}.
        \label{eq:rz_12_transmon}
\end{equation}

\subsubsection{Readout} 
The state discrimination process starts with applying a 700 ns duration rectangular pulse to the readout transmission line. In the experiment, we use a single-shot dispersive readout technique. During a readout calibration, we prepare qutrit in one of the states $\ket{0}$, $\ket{1}$ and $\ket{2}$ for $5 \cdot 10^4$ times each and measure the corresponding response trajectories $x_{i}(t)$ by a digitizer. 
The obtained trajectories are split into train and test sets. Then the train set is integrated in time with appropriate weight functions $F_0(t)$ and $F_1(t)$. In the current experiment, these functions are inspired by Gram–Schmidt orthogonalization process~\cite{krinner2022realizing} and defined as follows:
\begin{align}
    F_0(t) = \langle x_{1}^*(t)  - x_{0}^*(t)\rangle &, \\
    F_1(t) = \langle x_{2}^*(t)  - x_{0}^*(t)\rangle &- \nonumber\\
    - &\frac{\int F_0(t) (x_{2}(t)  -  x_{0}(t)) dt}{\int |F_0(t)| ^ 2 dt} F_0(t),
\end{align}
where $\langle . \rangle$ stands for the averaging over all trajectories and $^*$ denotes the complex conjugation operation.

The integration process projects each trajectory onto the weigh functions, which is equivalent to the quadratures calculation by the downsampling method
\begin{equation}
  \begin{gathered}
    I_{i} = \text{Pr}^0_{i} = \operatorname{Re}\left(\int F_0(t) x_{i}(t) dt \right),      \\
    Q_{i} = \text{Pr}^1_{i} = \operatorname{Re}\left(\int F_1(t) x_{i}(t) dt \right).
  \end{gathered}
\end{equation}
The calculated quadratures are generally represented as Gaussian clouds with a similar distribution in the IQ plane (see Fig.~\ref{fig:transmon_readout-calibartion}a).
{The obtained quadratures are classified by the standard machine-learning logistic regression method. We then use the test set to calculate a readout confusion matrix and fidelity as accuracy evaluation of the trained classification model. The resulted confusion matrix is shown in~Fig.~\ref{fig:transmon_readout-calibartion}b.
% The trained model for the test trajectories gives the readout confusion matrix shown in~Fig.~\ref{fig:transmon_readout-calibartion}b.
In our experiment, the average readout fidelity of a qutrit state classification is 87.6\%. 

%%%% FIGURE 9 %%%%
\begin{figure}
 \center{\includegraphics[width=0.48\textwidth]{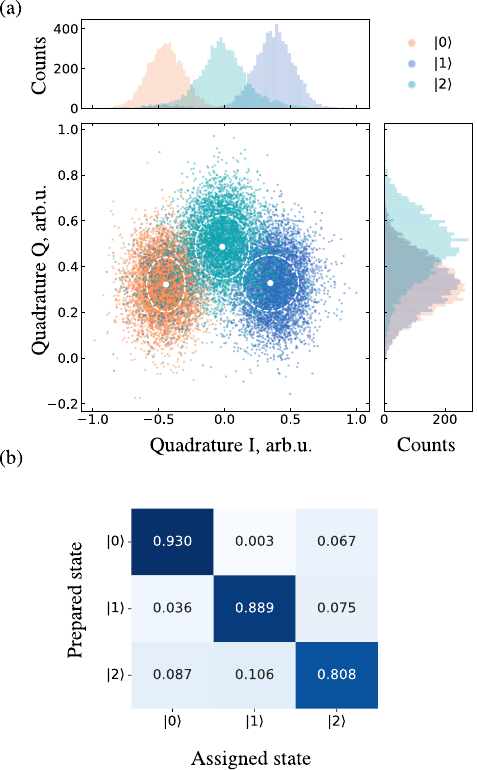}}
 \caption{(a) The readout calibration trajectories of qutrit states are presented in the $(I, Q)$ plane. Orange, blue, and aquamarine colored dots indicate measured Gaussian readout clouds corresponding to the $\ket{0}$, $\ket{1}$, and $\ket{2}$ states. The mean value and standard deviation of each cloud are denoted by white dots and dashed ellipses respectively. (b) The readout confusion matrix shows probability of readout declaration error. The average value of diagonal elements represents the total readout fidelity of the experiment.}
 \label{fig:transmon_readout-calibartion}
\end{figure}

\renewcommand{\bibname}{Reference}
\bibliographystyle{unsrtnat}
\bibliography{library}
\end{document}